\documentclass{iopconfser}
\usepackage{epsfig,cite,graphics}
\usepackage{marvosym} 

\usepackage[utf8]{inputenc}
\usepackage{natbib}

\usepackage[colorlinks=true, citecolor=blue, linkcolor=blue, urlcolor=blue]{hyperref}
\usepackage{multirow}
\usepackage{enumitem}

\newcommand{\aap}{A\&A}

\newcommand{\apjs}{The Astrophysical Journal Supplement Series}

\newcommand{\apjl}{ApJL}

\usepackage{fancyhdr}
\usepackage{lastpage}
\begin{document}

\title{Radar-Chart Analysis of Star-Formation Quenching Stages Across Circular Velocity Curve Classes in Nearby CALIFA Galaxies}

\author{Veselina Kalinova$^{1,2}$ and Dario Colombo$^{3}$}

\affil{$^1$Max Planck Institute for Radio Astronomy, Auf dem Hügel 69, 53121 Bonn, Germany}
\affil{$^2$Institute of Astronomy and National Astronomical Observatory,
Bulgarian Academy of Sciences, 72 Tsarigradsko Chaussee Blvd., 1784 Sofia, Bulgaria}
\affil{$^3$Argelander-Institut für Astronomie, University of Bonn, Auf dem Hügel 71, 53121 Bonn, Germany}

\email{kalinova@mpifr.de}

\begin{abstract}
We analyze the circular velocity curves of 215 non-active galaxies from the Calar Alto Legacy Integral Field spectroscopy Area survey. 
Using radar-chart analysis, we compare two classification schemes: the circular velocity curve classification
and the quenching classification.
We find a systematic progression from slow-rising to flat, round-peaked, and sharp-peaked curves that mirrors the transition from star-forming to fully retired systems. Star-forming galaxies are mainly slow-rising and flat, whereas fully retired galaxies are predominantly round-peaked and sharp-peaked, with intermediate stages showing a gradual shift. Overall, advanced quenching stages are associated with more centrally peaked curves and higher outer velocities (i.e., velocities along the flat part of the curve), indicating increasingly concentrated central mass distributions and more massive disks of the galaxies.
\end{abstract}

\section{Introduction}
Determining how galaxies evolve and what drives the cessation of star formation—commonly referred to as ``star-formation quenching"—remains one of the central challenges of modern astrophysics. Observations and diagnostic diagrams, such as color–mass and star-formation-rate–mass relations, suggest a bimodality in galaxy populations, distinguishing two main groups: star-forming and quiescent galaxies  (e.g., \citealt{Brinchmann2004}; \citealt{Baldry2004}; \citealt{Peng2010}). The global decline of the cosmic star-formation rate density since $z \sim 2$ (\citealt{Madau2014}) shows that quenching is a fundamental process in galaxy evolution (e.g., \citealt{Saintonge2017}, \citealt{Colombo2025a, Colombo2025b}). Nevertheless, the exact mechanisms responsible for suppressing star formation within individual galaxies remain debated.

Internal and external processes have been proposed as possible scenarios for star-formation quenching in nearby galaxies. Internal mechanisms include morphological quenching (\citealt{Martig2009}), bar-driven gas redistribution (\citealt{Romeo2015}), and feedback from active galactic nuclei (e.g., \citealt{Fabian2012}, \citealt{DiMatteo2005}). On the other hand,  environmental effects such as ram-pressure stripping and galaxy interactions may also contribute (e.g., \citealt{Abadi1999}). Observations suggest that quenching frequently proceeds inside-out  (e.g., \citealt{Perez2013}, \citealt{Gonzalez-Delgado2016}, \citealt{Belfiore2017}, \citealt{Sanchez2018}, \citealt{Kalinova2021}), pointing toward a close connection between galaxy morphology, dynamics, and suppression of star formation.

Galaxy dynamics provides direct information for the underlying mass distribution and gravitational potential of the studied systems (e.g., \citealt{Noordermeer2007b}). Circular velocity curves (CVCs) trace the radial distribution of total mass and contain information about the relative contributions of bulge, disk, and dark matter halo components (\citealt{Rubin1980}, \citealt{Bosma1981}). The shape and amplitude of the CVC therefore reflect the structural configuration and assembly history of a galaxy (e.g., \citealt{Avila-Reese2002}). A dynamical classification introduced by \cite{Kalinova2017b} reflects the variations of galaxy CVCs (according to their amplitudes and shapes) and defines four main CVC classes: slow-rising (SR), flat (FL), round-peaked (RP), and sharp-peaked (SP), corresponding to different mass distributions.

Complementing this framework, \cite{Kalinova2021} defined a spatially resolved quenching classification based on the ionised gas distribution traced by the H$\alpha$ equivalent width map. This scheme distinguishes six star-formation quenching stages (QS), ranging from star-forming galaxies to fully retired systems: star-forming (SF), quiescent-nuclear-ring (QnR), centrally quiescent (cQ), mixed (MX), nearly retired (nR), and fully retired (fR), and captures the progressive suppression of star formation across galactic disks. The classification provides a physically motivated evolutionary sequence that links spatially resolved star-formation properties with global ones.

Furthermore, \cite{Kalinova2022} analyzed 215 nearby non-active galaxies to explore whether a galaxy’s inner gravitational potential, traced by its circular velocity curve (CVC), is linked to local star-formation quenching. Using pixel-by-pixel measurements, they found that regions with higher circular velocities tend to be quenched, while lower-velocity regions are mostly star-forming. They also find that both the amplitude and shape of the CVC are closely connected to  suppression of star formation, supporting dynamical or morphological quenching as an important driver of galaxy evolution (e.g., \citealt{Martig2009}, \citealt{Gensior2020}).

The main goal of this study is to extend these previous works and further investigate the link between galaxy dynamics and star-formation activity in nearby galaxies by combining the CVC classification of \cite{Kalinova2017b} with the quenching classification of \cite{Kalinova2021} in a more quantitative way. Unlike \cite{Kalinova2022}, which rely on pixel-by-pixel analyses, we quantify the relationship between the two classifications by treating them as discrete categories, enabling a more statistical assessment of their association. To do this, we employ a radar-chart analysis that visualizes the relative distribution of CVC classes within each QS, and investigate how tight is the link between dynamics and star-formation quenching per given class. Studying such correlation, we will be able to set an observational constraints in the future on the interplay between mass distribution, structure of the galaxy, and the mechanisms responsible for quenching in the local Universe.

\begin{figure*}
\includegraphics[width=1\textwidth]{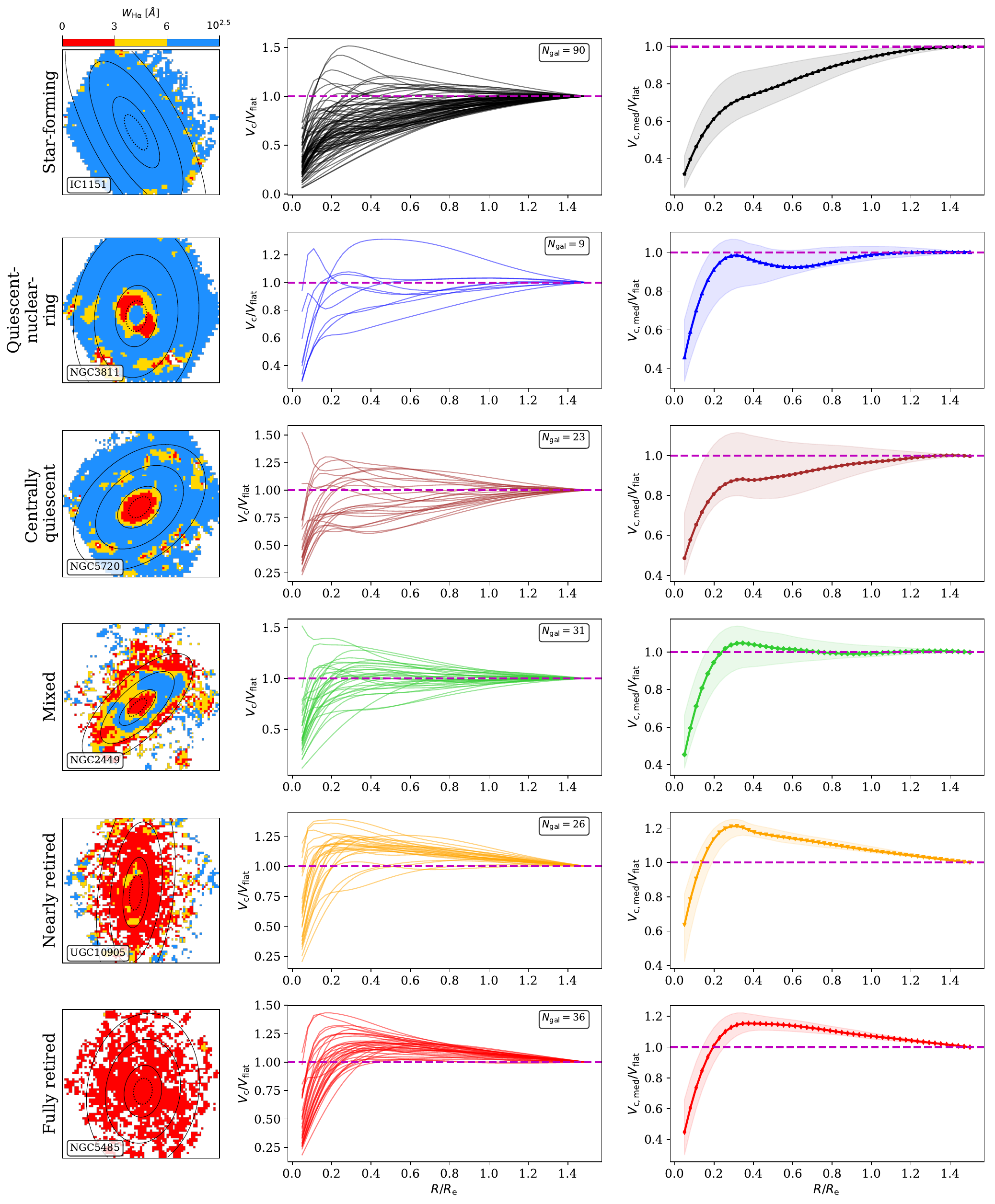}
\caption{\footnotesize{CVCs of galaxies across the quenching sequence defined in \cite{Kalinova2021}.
\emph{First column} (from top to bottom): Star-formation quenching classification of our sample into star-forming, quiescent nuclear ring, centrally quiescent, nearly retired, and fully retired galaxies. Note that this column is based on the same data as Fig. 1 in \cite{Kalinova2022}, and we re-plot it here for comparison and representative purposes.
\emph{Second column:} CVCs normalised by the asymptotic circular velocities of the galaxies for each QS.
\emph{Third column:} Median circular velocity curve for all galaxies within a given QS; the shaded regions indicate the interquartile range (25th–75th percentiles) of the sample. } }
\label{fig:cvc-qs}
\end{figure*}

\begin{figure*}
\includegraphics[width=1\textwidth]{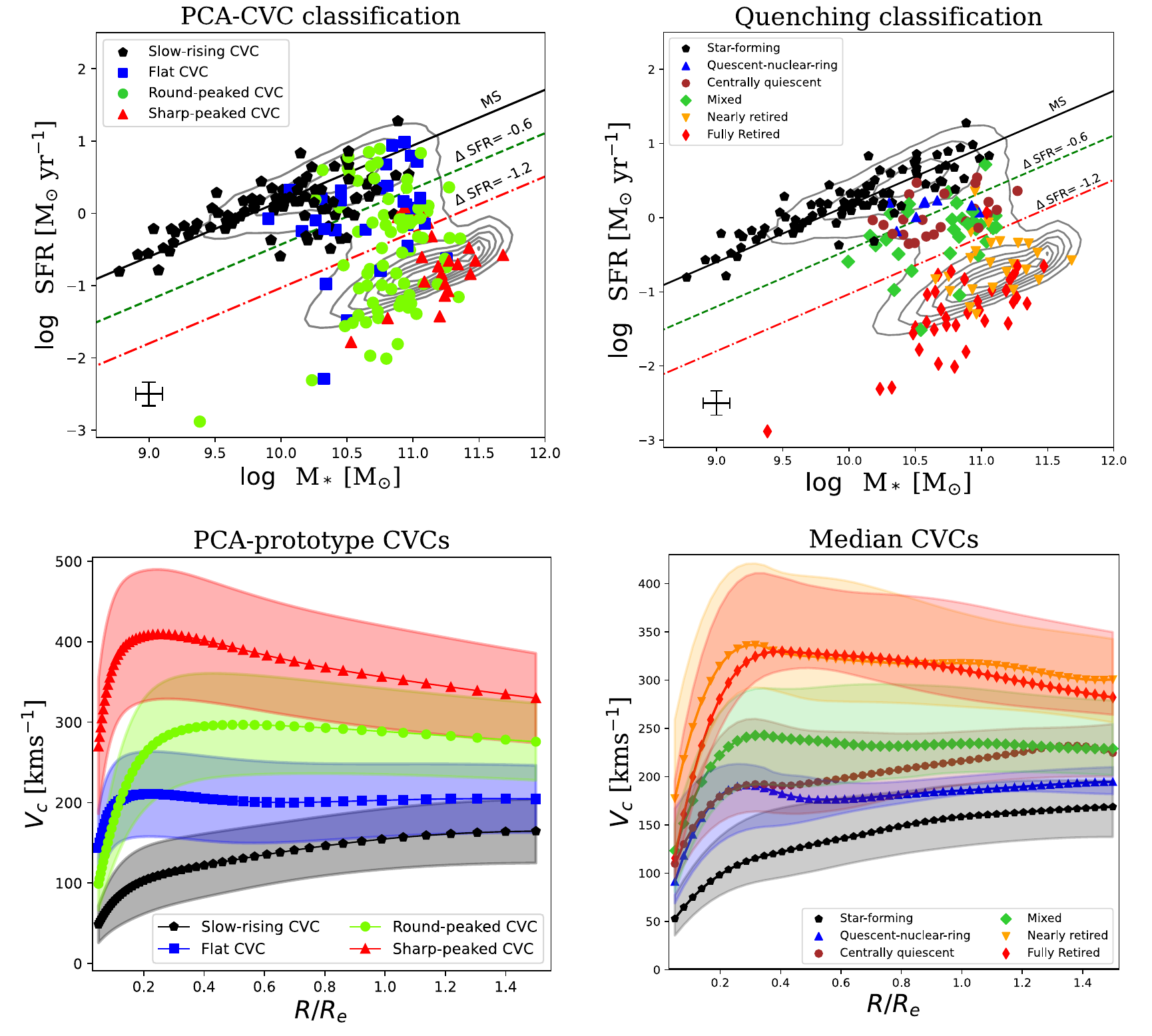}
\caption{\footnotesize{\emph{Top row:} $SFR–M_\ast$ diagram shown for the CVC classification (left) and the quenching classification (right). The thick black line, the dashed green line, and the red dash–dotted line indicate the Main Sequence of star formation, the onset of the Green Valley, and the onset of the Red Sequence, respectively. The Green Valley region is defined by the thresholds $-1.2 < \Delta \mathrm{SFR} < -0.6$, following \cite{Bluck2016}. 
\emph{Bottom row:} Prototype CVCs for each CVC class (left), derived via PCA analysis and $k$-means clustering technique following \cite{Kalinova2017b}, with shaded uncertainty bands. The right panel shows the median CVCs for each QS, with the corresponding uncertainties indicated by the shaded regions. Note that the data shown in the four panels of this figure have been published previously (top-left panel: \citealt{Kalinova2026a}; top-right panel: \citealt{Kalinova2021}; bottom-left panel: \citealt{Kalinova2017b}; bottom-right panel: \citealt{Kalinova2022}). We re-plot them here for completeness and to provide representative context, as they are directly related to the subsequent analysis presented in this paper.}}
\label{fig:SFR-mass}
\end{figure*}

\section{Sample and Data}
We analyse 215 non-active nearby galaxies from CALIFA survey (Calar Alto Legacy Integral Field spectroscopy Area; \citealt{Sanchez2012}) across Hubble sequence with various morphologies from elliptical to late-type spirals (E1$-$Sdm) and large range of stellar masses (from $\sim 10^{8}$ M$_{\odot}$ to $\sim$ $10^{11}$ M$_{\odot}$). The current sample is drawn from the sample of 238 galaxies in \cite{Kalinova2017b} and \cite{Kalinova2021}. Our sample is representative of the CALIFA mother sample as well as of the galaxy population in the nearby Universe (\citealt{Walcher2014}; \citealt{Kalinova2017b}).

We adopt the stellar mass ($M_\ast$), star formation rate ($SFR$) values, and QS of our galaxies from \cite{Kalinova2021}, while the CVC profile of individual galaxy and their CVC classes  are taken from \cite{Kalinova2017b}.

\section{Method \& Results}
\label{S:method-results}
To explore the link between gravitational potential and star-formation quenching of our sample of galaxies, we construct several diagnostic plots in comparison. 
Visual representations of the CVCs of the galaxies, grouped across QSs, are presented in Figure~\ref{fig:cvc-qs}. In the first-column panels, the main QSs are shown alongside the corresponding CVCs of our sample, normalized by the asymptotic velocity in the second-column panels. In the third-column panels, the median CVC of all galaxies in each QS is calculated, together with the corresponding uncertainty band. Already from this combined plot, we can identify a clear trend - galaxies from the star-forming QSs (SF, QnR, cQ) tend to have slow-rising to flat CVCs, while galaxies towards QSs (MX, nR, fR) tend to have more peaked CVCs.

\begin{figure*}[h]
\includegraphics[width=1\textwidth]{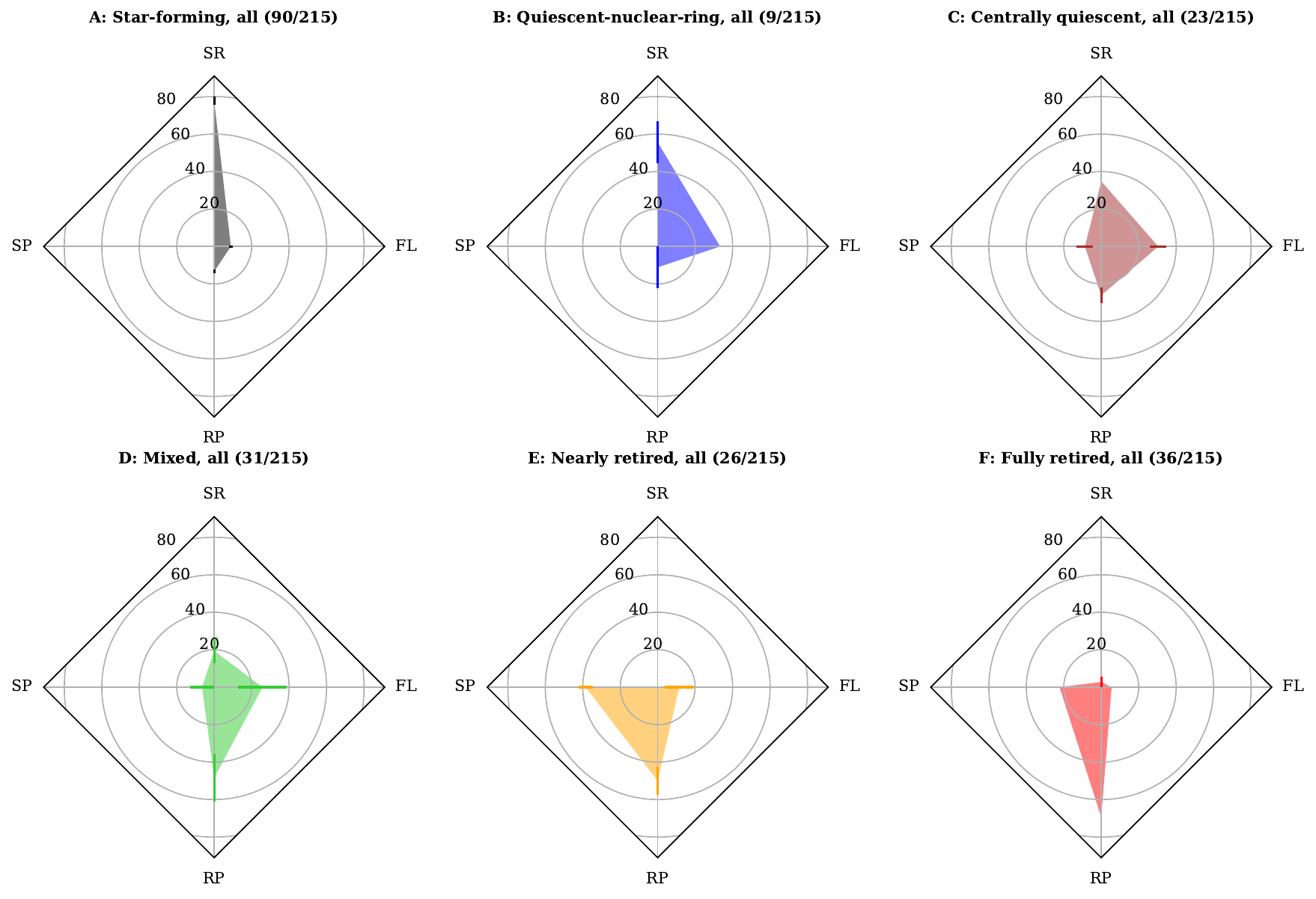}
\caption{\footnotesize{Radar-chart analysis of star-formation QSs across CVC classes for our sample of 215 CALIFA galaxies. The charts highlight a gradual increase in both the central peak and the overall amplitude of the CVCs along the quenching sequence. The radial extension along each axis reflects the uncertainty for each variable. Moving clockwise around the radar chart, the CVC steepness and amplitude increases.  }}
\label{fig:radar-chart-individual}
\end{figure*}

\begin{figure*}[h]
\centering
\includegraphics[width=0.7\textwidth]{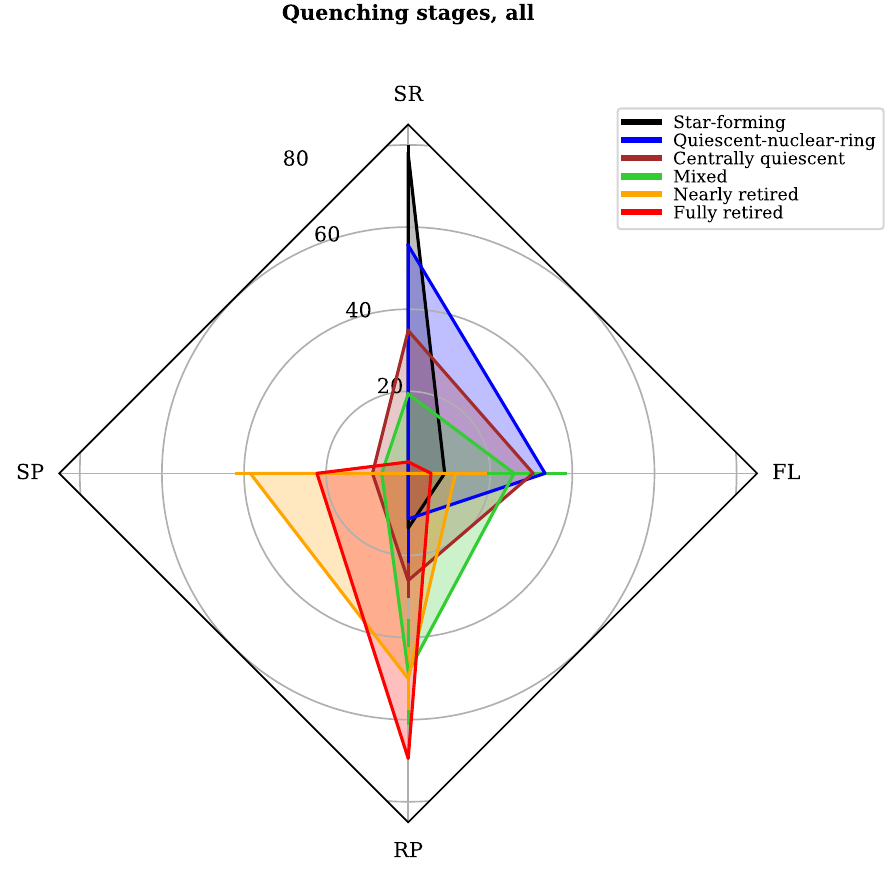}
\caption{\footnotesize{Combined radar chart of all QSs across CVC classes. This figure, similar to Fig. \ref{fig:radar-chart-individual}, emphasizes the gradual increase in both the central peak and the overall amplitude of the CVCs along the quenching sequence. The radial extension along each axis reflects the uncertainty for each variable. Moving clockwise around the radar chart, the CVC steepness and amplitude increases. }}
\label{fig:radar-chart-combined}
\end{figure*}

Furthermore, Figure~\ref{fig:SFR-mass} shows the prototype/median curves with their associated uncertainty bands for each CVC class (bottom-left panel) and each QS (bottom-right panel). The same colour-coding highlights the location of each galaxy group in the $SFR-M_\ast$ plane (top-row panels). The QSs are well separated across the Main Sequence of star formation, the Green Valley, and the Red Sequence. The three CVC classes—SR, FL, and SP—follow a similar trend, with the exception of RP-CVC galaxies, which spread from the Main Sequence to the Red Sequence. It is possible that this effect arises due to the lower number of classes in the CVC classification compared to the number of stages in the quenching classification. In the RP category, MX and fR galaxies appear to be treated as a single combined population.

To examine quantitively how galaxies are distributed across CVC classes at each QS, we apply a ``radar-chart analysis''. Figure \ref{fig:radar-chart-individual} presents the individual radar chart per QS, while Fig.  \ref{fig:radar-chart-combined} combine the radar charts of all QSs on one plot. 
Radar chart (also known as spider or polar diagrams;  e.g., \citealt{Chambers1983}, \citealt{Wilkinson2005}) is a graphical method that visualise multivariate data on a two-dimensional plane. It compares multiple datasets (in our case two: QS and CVC class) by forming polygons.
Each variable is represented by an axis originating from a common central point. The axes are arranged radially similar to the spokes on a wheel. The value (percentage) of each variable is plotted along its corresponding axis. The points are connected, forming a shape of polygon, where larger polygon's surface indicates higher values (percentage in our case).

Figure \ref{fig:radar-chart-individual} and \ref{fig:radar-chart-combined} represent the QSs as fractions of the four CVC classes. For  each CVCs class is given an axis, spaced equally and arrange radially around the centre. Since the CVC classes are four, their axes are separated by 90 degrees. The CVC classes (SR, FL, RP, and SP) are located in one of the four pivots of the radar chart for each QS, where in clockwise direction the  CVC steepness and amplitude increases. The circles in the square of the radar chart correspond to the percentage ratio of the CVC classes (i.e.,  20\%, 40 \%, 60\% and 80\% from the centre to the edges of the chart)  in respect of the total galaxies' number of each QS. The radial extension along each axis of the radar charts in Fig. \ref{fig:radar-chart-individual} and \ref{fig:radar-chart-combined} reflects the uncertainties, which are estimated as the number of the unsure classes due to a borderline cases or poor data (as described in Sec. 3.3 in \citealt{Kalinova2021} versus the total number of the galaxies per QS).

Table~\ref{tab:tab1} presents the fractional distribution of CVC classes across the different QSs, expressed as percentages. These fractions correspond to the data visualized in the radar charts shown in Figs.~\ref{fig:radar-chart-individual} and \ref{fig:radar-chart-combined}. Each row lists a QS category — SF, QnR, cQ, MX, nR, and fR — while the columns indicate the percentage of galaxies in each CVC class: SR, FL, RP, and SP.

\begin{table}[h]
\centering
\caption{Fractional distribution of CVC classes across QSs, expressed as percentages in the radar chart shown in Fig. \ref{fig:radar-chart-individual} and \ref{fig:radar-chart-combined}.}
\begin{tabular}{c|cccc}
\hline
QS & SR & FL & RP & SP \\
\hline
SF & 73.5 & 27.3 & 18.3 & 0.0 \\
QnR & 7.1 & 12.1 & 1.2 & 0.0 \\
cQ & 8.2 & 21.2 & 7.3 & 8.0 \\
MX & 9.2 & 24.2 & 22.0 & 8.0 \\
nR & 1.0 & 9.1 & 20.7 & 52.0 \\
fR & 1.0 & 6.1 & 30.5 & 32.0 \\
\hline
\end{tabular}
\label{tab:tab1}
\end{table}

Our results show that SF galaxies are predominantly associated with SR CVCs, whereas nR and fR galaxies exhibit higher fractions of RP and SP CVCs. The intermediate QSs (QnR, cQ, and MX) display a mixture of CVC classes. More quantitatively, galaxies with SR CVCs are predominantly represented by SF systems (73.5\%). FL CVC galaxies are mainly associated with SF (27.3\%), MX (24.2\%), and cQ (21.2\%) systems. RP CVC galaxies are primarily found among fR (30.5\%), MX (22.0 \%), and nR (20.7 \%) systems, whereas SP CVC galaxies are mostly associated with nR (52.0\%) and fR (32.0\%) systems.

In particular, as galaxies evolve from star-forming to retired stages, they progressively transit from CVC classes with shallow slopes and low amplitudes  to classes whose CVCs exhibit a more pronounced central peak and higher overall amplitudes. This trend suggests that the quenching process is linked to the growth or the increasing dynamical dominance of central structures—such as bulges or bars—and to a deepening of the gravitational potential.

Interestingly, nR galaxies that still retain residual star-formation activity in their disks show the largest fraction of SP CVCs, which are associated with the highest velocity amplitudes. This may indicate that even galaxies close to full quenching can maintain dynamically massive disk components, highlighting the complex interplay between mass distribution and star-formation activity.

Because the shape of the CVC traces the underlying gravitational potential and mass distribution, the systematic shift among CVC classes along the quenching sequence supports a scenario in which structural and dynamical evolution are closely connected to the quenching of star formation.

\section{Conclusions}
In this study, we examine 215 non-active galaxies from the CALIFA survey. We apply a radar-chart analysis to investigate the link between CVC classes and QSs. We find a strong connection between the gravitational potential of galaxies, as traced by their CVCs, and their QS. In particular, we quantify the distribution of galaxies across both the CVC classification and the quenching sequence.

Our results suggest that the gravitational potential of different galaxy components may play a central role in star-formation quenching process and, consequently, in galaxy evolution. Future work will further explore this connection in a greater detail.

\section*{Acknowledgements}
\footnotesize{DC gratefully acknowledges the Collaborative Research
Center 1601 (SFB 1601 sub-project B3) funded by the Deutsche Forschungsgemeinschaft (DFG, German Research Foundation) –
500700252.
In this study, we made use of the data of the first legacy survey, 
the Calar Alto Legacy Integral Field Area (CALIFA) survey, 
based on observations made at the Centro Astron\'omico
Hispano Alem\'an (CAHA) at Calar Alto, operated jointly by the Max Planck-Institut
f\"ur Astronomie and the Instituto de Astrof\'isica de Andaluc\'ia
(CSIC).
This research made use of the open-source python packages as \texttt{Astropy} (\citealt{Price2018astropy}), \texttt{SciPy} (\citealt{SciPy2020}), \texttt{NumPy} (\citealt{Harris2020NumPy}), and \texttt{Matplotlib} (\citealt{Hunter2007matplotlib}). }

\bibliographystyle{aasjournal}
\footnotesize{\bibliography{BIB_v10}}

\end{document}